%% file: paper.tex
\documentclass[english]{revtex4}
\usepackage[T1]{fontenc}
\usepackage[latin1]{inputenc}
\usepackage{amsmath}
\usepackage{graphicx}
\usepackage{amssymb}

\makeatletter
\newcommand{\linea}{\noindent\rule{1.0\textwidth}{1pt}}%

\newenvironment{algo}{\medskip\noindent\begin{minipage}{1.0\textwidth}\linea\begin{sf}\\}{\end{sf}\linea\end{minipage}\bigskip}

\usepackage{color}

\usepackage{babel}
\makeatother
\begin{document}

\title{Survey propagation: an algorithm for satisfiability}
\author{A.
Braunstein$^{(1,3)}$, M. M\'ezard$^{(2)}$, R. Zecchina$^{(3)}$}
\affiliation{ $^{1}$International School for Advanced Studies (SISSA),\\ via Beirut 9, 34100 Trieste, Italy \\
 $^{2}$Laboratoire de Physique Th\'eorique et Mod\`eles
Statistiques, CNRS and Universit\'e Paris Sud, B\^at. 100, 91405
Orsay
{\sc cedex}, France\\
$^1$The Abdus Salam International Centre for Theoretical Physics (ICTP),
\\ Str. Costiera 11, 34100 Trieste, Italy
\\
}

\date{\today}

\begin{abstract}

We study the satisfiability of randomly generated formulas formed by
$M$ clauses of exactly $K$ literals over $N$ Boolean variables.  For a
given value of $N$ the problem is known to be most difficult when
$\alpha=M/N$ is close to the experimental threshold $\alpha_c$ separating
the region where almost all formulas are SAT from the region where all
formulas are UNSAT.  Recent results from a statistical physics
analysis suggest that the difficulty is related to the existence of a
clustering phenomenon of the solutions when $\alpha$ is close to (but
smaller than) $\alpha_c$. We introduce a new type of message passing
algorithm which allows to find efficiently a satisfiable assignment of
the variables in this difficult region.  This algorithm is iterative
and composed of two main parts. The first is a message-passing
procedure which generalizes the usual methods like Sum-Product or
Belief Propagation: it passes messages that are surveys over clusters
of the ordinary messages. The second part uses the detailed
probabilistic information obtained from the surveys in order to fix
variables and simplify the problem. Eventually, the simplified problem
that remains is solved by a conventional heuristic.
\end{abstract}

\maketitle
\section{Introduction}

The satisfiability problem is the archetype of combinatorial
optimization problems which are well known to be intractable in the
worst case. However, experimental studies show that many instances of
satisfiability are surprisingly easy, even for naive heuristic
algorithms.  In an attempt to get a better understanding of which
instances are easy or hard to solve, a lot of efforts have focused in
recent years on the 'random K-sat' problem\cite{Cook_review}.
Instances of this problem are generated by considering $N$ variables
and $M=\alpha N$ clauses, where each clause contains exactly $K$
distinct variables, and is picked up with uniform probability
distribution from the set of $\left(\begin{array}{c}N \\ K
\end{array}\right) 2^K$ possible clauses. For a given value of
$\alpha$, the probability $P_N(\alpha)$ that a randomly generated
instance is SAT is a decreasing function, with $P_N(0)=1$ and
$\lim_{\alpha \to \infty}P_N(\alpha=0)$, which has been shown to
approach, as $N$ increases, a step function characteristic of a
zero-one law \cite{friedgut}, or a `phase transition'. It is
convenient to identify a crossover regime between 'SAT' and 'UNSAT'
regimes using the value $\alpha_c(N)$ of the number of constraints per
variable where $P_N(\alpha_c(N))=1/2$.  From numerical simulations,
$\alpha_c(N)$ is supposed to converge, in the large $N$ limit, to a
value around $\alpha_c \simeq 4.27$
\cite{TCS_issue,KirkSel,Crawford,MZKST}, but this convergence has not
yet been established rigorously. Interestingly, the performance of
algorithms is found to be much worse around this value of
$\alpha=4.27$: randomly generated instances with  $\alpha$ near to
the phase transition are particularly difficult to solve.

Rigorous lower and upper bounds have been found for this conjectured
satisfiability threshold: it has been established that $\lim_{N \to
\infty} P_N(\alpha)=1$ for $\alpha<\alpha_{lb}$ and $\lim_{N \to
\infty} P_N(\alpha)=0$ for $\alpha>\alpha_{ub}$. The present best
bounds for the case of the random 3-SAT problem (with $K=3$) are
$\alpha_{ub}=4.506$ (from \cite{dubois}, using the first moment
method) and $\alpha_{lb}=3.42$ (from \cite{Kirousis_lb}, using
algorithmic analysis). Note also the interesting algorithm-independent
upper bound found in \cite{moore_achl,achl_perez} using the second
moment method, which becomes better for larger values of $K$.

Recently, some elaborate statistical physics methods have been brought
to bear on the random satisfiability problem.  These non-rigorous
analytical calculations have put forward some interesting conjectures
about what happens in the solution space of the problem as this
threshold is approached\cite{MEPAZE,MZ_pre} (see also previous work in
\cite{MonZec,Biroli}). They suggest the following overall picture,
which should hold for a generic sample of the random satisfiability
problem, in the limit $N\to\infty$, with $\alpha$ fixed.
\begin{enumerate}
\item
There exists a SAT-UNSAT phase transition at a critical value
$\alpha_c$ which can be computed by solving some (complicated)
integral equation; for $K=3$, one gets $\alpha_c\simeq 4.267$

\item 
There exists a second threshold $\alpha_{clust}$ separating two phases
which are both `SAT' (in each of them there exists a satisfiable
assignment with probability 1), but with very different geometric
structures:

\item
For $\alpha<\alpha_{clust}$, a generic problem has many solutions,
which tend to form one giant ``cluster''; the set of all satisfying
assignments forms a connected cluster in which it is possible to find
a path between two solutions that requires short steps only (each pair
of consecutive assignments in the path are close together in Hamming
distance). In this regime, local search algorithms and other simple
heuristics can relatively easily find a solution.  This region is
called the 'easy-SAT' region

\item 
For $\alpha_{clust}<\alpha<\alpha_c$, there exists 
a 'hard SAT' phase where the solution space breaks up
into many smaller clusters.  Solutions in separate clusters are
generally far apart: it is not possible to transform a SAT assignment
in one cluster into another one in a different  cluster by changing only
a finite number of variables.  Because of this clustering effect,
local search algorithms tend to have a very slow convergence
when applied to large $N$ instances.

\end{enumerate}

So far,  the analytic method used in the most recent statistical physics
analysis, named the cavity method\cite{Bethe_cav},  is non 
rigorous, and turning this type of approach into a rigorous 
theory is an open subject of current research\cite{talag,FraLeo}.
Note however that, in the simpler case of the random K-XOR-SAT, the
validity of this statistical physics analysis can be confirmed by
rigorous studies \cite{xorsat1,xorsat2}, and the above clustering
conjecture has been fully confirmed.

Interestingly, the statistical physics analysis suggests a new
efficient heuristic algorithm for finding SAT assignments in the hard
SAT phase, which has been put forward by two of us in \cite{MZ_pre}.
The aim of this paper is to provide a detailed self-contained
description of this algorithm, which does not rely on the statistical
physics background. We shall limit the description to the regime where
solutions exist, the so called SAT phase; some modification of the
algorithm allows to address the optimization problem of minimizing the
number of violated constraints in the UNSAT phase, but it will not be
discussed here.

The basic building block of the algorithm, called survey propagation
(SP), is a message passing procedure which resembles in some respect
the iterative algorithm known as belief propagation (BP), but with
some crucial differences which will be described.  BP is a generic
algorithm for computing marginal probability distributions in problems
defined on factor graphs, which has been very useful in the context of
error correcting codes \cite{Gallager} and Bayesian networks
\cite{pearl}.

While in simple limits we are able to give some rigorous results
together with an explicit comparison with the belief propagation
procedures, in general there exists no rigorous proof of convergence
of the algorithm.  However, we provide clear numerical evidence of its
performance over benchmarks problems which appear to be far larger
than those which can be handled by present state-of-the-art
algorithms.

The paper is organized as follows: Sect. \ref{sect_fg} describes the
satisfiability problem and its graphical representation in terms of a
factor graph.
Sect. \ref{sect_WP} explains two message passing algorithms, namely
warning propagation (WP) and belief propagation (BP).  Both are exact
for tree factor graphs. Even if they are typically unable to find a
solution for random SAT in the ``interesting'' hard-SAT region, they
are shown here because they are in some sense the basic building
blocks of our survey propagation algorithm. Sect. \ref{wptosp}
explains the survey propagation algorithm itself, a decimation
procedure based on it, the `survey inspired decimation' (SID), and the
numerical results. In Sect.\ref{heuristic} we give some heuristic
arguments from statistical physics which may help the reader to
understand where the SP algorithm comes from. Sect.~\ref{sect_concl}
contains a few general comments.

\section{The SAT problem and its factor graph representation}
\label{sect_fg}
We consider a satisfiability problem consisting of $N$ Boolean
variables $\{ x_i \in \{0,1\} \}$ (where $\{0,1 \} \equiv \{F,T\}$), with $i
\in \{ 1,...,N \}$, with $M$ constraints.  Each
constraint is a clause, which is the logical OR of the variables or of
their negations. A clause $a$ is characterized by the set of variables
$i_1,...,i_K$ which it contains, and the list of those which are
negated, which can be characterized by a set of $K$ numbers $J^a_{i_r}
\in \{ \pm 1\}$ as follows.  The clause is written as
\begin{equation}
 \left ( z_{i_1} \vee ...\vee z_{i_r} \vee ... \vee z_{i_K} \right)
\end{equation}
where $z_{i_r}=x_{i_r}$ if $J^a_{i_r}=-1$ and $z_{i_r}=\bar x_{i_r}$
if $J^a_{i_r}=1$ (note that a positive literal is represented by a
negative $J$). The problem is to find whether there exists an
assignment of the $x_i \in \{ 0,1\}$ which is such that all the $M$
clauses are true.  We define the total cost $C$ of a configuration
${\bf x}=(x_1,...,x_N)$ as the number of violated clauses.

In what follows we shall adopt the factor graph
representation~\cite{factor_graph} of the SAT problem. This
representation is convenient because it provides an easy graphical
description to the message passing procedures which we shall
develop. It also applies to a wide variety of different combinatorial
problems, thereby providing a unified notation.

\begin{figure}
\centering
\includegraphics[width=7.cm]{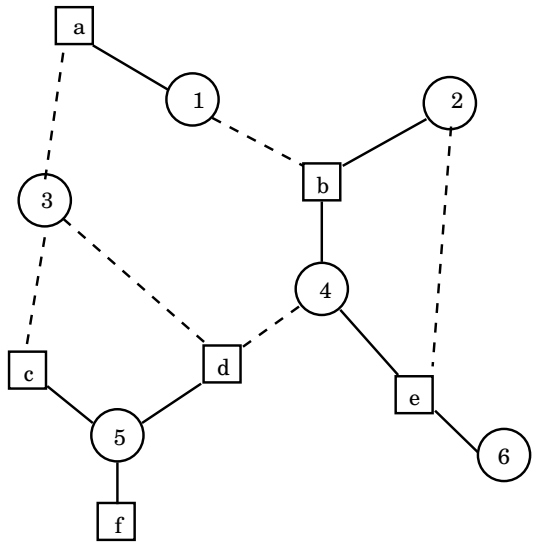}
\caption{
An example of a factor graph with $6$ variable nodes $i=1,..,6$ and
$6$ function nodes $a,b,c,d,e,f$.
The formula which is encoded is: $F= ( x_1 \vee \bar x_3 )
\wedge ( \bar x_1 \vee   x_2 \vee x_4)\wedge (\bar x_3 \vee x_5) \wedge 
(\bar x_3 \vee \bar  x_4\vee  x_5)\wedge (\bar x_2 \vee x_4 \vee x_6)
\wedge (x_5)
$}
\label{fig_fgdef}
\end{figure}

The SAT problem can be represented graphically as follows (see
fig.\ref{fig_fgdef}). Each of the $N$ variables is associated to a
vertex in the graph, called a ``variable node'' (circles in the
graphical representation), and each of the $M$ clauses is associated
to another type of vertex in the graph, called a ``function node''
(squares in the graphical representation). A function node $a$ is
connected to a variable node $i$ by an edge whenever the variable
$x_i$ (or its negation) appears in the clause $a$.  In the graphical
representation, we use a full line between $a$ and $i$ whenever the
variable appearing in the clause is $x_i$ (i.e. $J^a_i=-1$), a dashed
line whenever the variable appearing in the clause is $\bar x_i$
(i.e. $J^a_i=1$). Variable nodes compose the set $X$ ($|X|=N$) and
function nodes the set $A$ ($|A|=M$).

In summary, each SAT problem can be described by a bipartite graph, $
G=\left( X \cup A ; E= X \times A\right)$ where $E$ is the edge set,
and by the set of ``couplings'' $\{ J_a^{i} \}$ needed to define each
function node.  For the K-SAT problem where each clause contains $K$
variables, the degree of all the function nodes is $K$.

\begin{figure}
\centering
\includegraphics[width=7.cm]{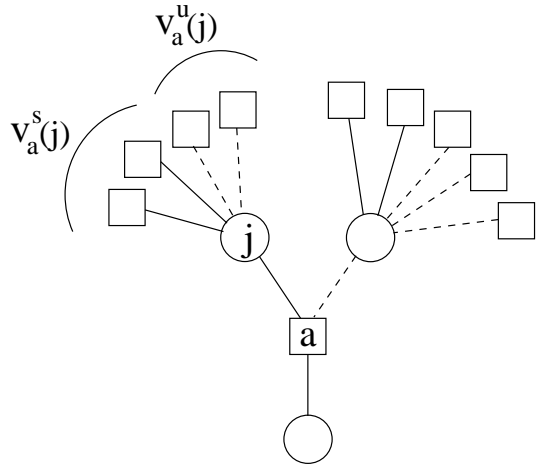}
\caption{A function node $a$ of the factor graph with the $V^u_a(j)$
and $V^s_a(j)$ sets relative to node $j$.}
\label{fig_iter}
\end{figure}

Throughout this paper, the variable nodes indices are taken in
$i,j,k,...$, while the function nodes indices are taken in
$a,b,c,...$. For every variable node $i$, we denote by $V(i)$ the
set of function nodes $a$ to which it is connected by an edge, by
$n_i=|V(i)|$ the degree of the node, by $V_+(i)$ the subset
of $V(i)$ consisting of function nodes $a$ where the variable
appears un-negated (the edge $(a,i)$ is a full line), and by
$V_-(i)$ the complementary subset of $V(i)$ consisting of function
nodes $a$ where the variable appears negated (the edge $(a,i)$ is a
dashed line). $V(i)\setminus b$ denotes the set V(i) without a node $b$.
Similarly, for each function node $a$, we denote by
$V(a)=V_+(a)\cup V_-(a) $ the set of neighboring variable nodes,
decomposed according to the type of edge connecting $a$ and $i$,
and by $n_a$ the degree. Given a function node $a$ and a variable node $j$,
connected by an edge, it is also convenient to define the two sets:
$V^u_a(j)$ and $V^s_a(j)$, where the indices $s$ and $u$
respectively refer to the neighbors which tend to make variable
$j$ satisfy or unsatisfy the clause $a$, defined as (see fig.\ref{fig_iter}):
\begin{eqnarray}
{\mbox{if}} \; && J_j^a=1 \; : \; \ V^u_a(j)= V_+(j) \ \ ; \ \
V^s_a(j)= V_-(j)\setminus a\\ {\mbox {if}}\; && J_j^a=-1: \; \
V^u_a(j)= V_-(j)\ \ ; \ \ V^s_a(j)= V_+(j)\setminus a
\end{eqnarray}

The same kind of factor graph representation can be used 
for other constraint satisfaction problems,
where each function node $a$ defines an arbitrary function over the
set $X_a \subset X$ of variable nodes to which is connected, and could
also involve hidden variables.

\section{The message passing solution of SAT on a tree}
\label{sect_WP}
In the special case in which the factor graph of a SAT problem is a
tree (we shall call it a tree-problem), the satisfiability problem can
be easily solved by many methods. Here we shall describe two message
passing algorithms.  The first one, called warning propagation (WP),
determines whether a tree-problem is SAT or not; if it is SAT, WP
finds one satisfiable assignment.  The second algorithm, called belief
propagation (BP), computes the number of satisfiable assignments, as
well as the fraction of these assignments where a given variable is
set to true. These algorithms are exact for tree-problems, but they
can be used as heuristic in general problems, and we first give their
general definition, which does not rely on the tree-like structure of
the factor graph.

\subsection{Warning propagation}
The basic elementary message passed from one function node $a$ to a
variable $i$ (connected by an edge) is a Boolean number $u_{a \to i}
\in\{ 0,1\}$ called a `warning'.

The update rule is defined as follows. Given a function node $a$ and
one of its variables nodes $i$, the warning $u_{a \to i}$ is
determined from the warnings $u_{b \to j}$ arriving on all the
variables $j \in V(a)\setminus i$ according to:
\begin{equation}
u_{a \to i}= \prod_{j \in V(a)\setminus i} \theta\left(
 - J^a_j \left(\sum_{b \in V(j)\setminus a} J^b_j u_{b \to j} \right)
\right) \ ,
\label{udef}
\end{equation}
where $\theta (x) =0$ if $x \leq 0$ and $\theta (x) =1$ if $x > 0$.
This update rule is used sequentially, resulting in the following
algorithm:

\begin{algo}
\noindent\textbf{WP} algorithm

\noindent INPUT: the factor graph of a Boolean formula in conjunctive normal
form; a maximal number of iterations $t_{max}$

\noindent OUTPUT: UN-CONVERGED if WP has not converged after $t_{max}$
sweeps. If it has converged: the set of all warnings $u_{a \to
i}^*$.

\linea

\begin{itemize}
\item[0.] At time $t=0$: For every edge $a \to i$ of the factor graph,
randomly initialize the warnings $u_{a \to i}(t=0) \in \{ 0,1\}$, 
e.g. with probability $1/2$.
\item[1.] For  $t= 1$ to $t=t_{max}$:
\begin{itemize}
\item[1.1] sweep the set of edges in a random order, and update
 sequentially the warnings on all the edges of the graph,
 generating the values $u_{a \to i}(t)$, using subroutine WP-UPDATE.
\item[1.2]
If $u_{a \to i}(t)=u_{a \to i}(t-1)$ on all the edges, the iteration
has converged and generated $u_{a \to i}^*=u_{a \to i}(t)$: go to 2.
\end{itemize}
\item[2.] If  $t=t_{max}$ return UN-CONVERGED. If $t<t_{max}$
return the set of  fixed point warnings $u_{a \to i}^*=u_{a \to i}(t)$
\end{itemize}

\noindent\linea

Subroutine WP-UPDATE$(u_{a \to i})$

\noindent
INPUT: Set of all warnings arriving onto each variable node $j
\in V(a) \setminus i$

\noindent
OUTPUT: new value for the warning $u_{a \to i}$.

\begin{itemize}
\item[1] For every $j \in V(a) \setminus i$, compute the cavity field
$ h_{j \to a}=\left(\sum_{b \in V_+(j) \setminus a} u_{b \to j}\right)
- \left(\sum_{b \in V_-(j)\setminus a}u_{b \to j}\right)$
(If $V(j) \setminus a$ is empty, then $ h_{j \to a}=0$).
\item[2] Using these cavity fields $h_{j \to a}$, compute the warning
$u_{a \to i}= \prod_{j \in V(a)\setminus i} \theta\left(h_{j \to a} J^a_j\right)$
(If $ V(a)\setminus i$ is empty, then $u_{a \to i}=1$).
\end{itemize}

\end{algo}

\begin{figure}
\centering
\includegraphics[width=8.cm]{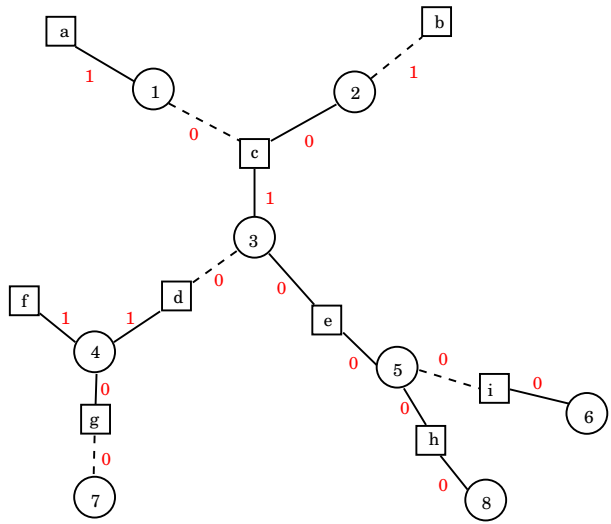}
\caption{
 An example of result obtained by the WP algorithm on a tree-problem
with $N=8$ variables and $M=9$ clauses. The number  on each edge of the graph
is the value of the corresponding warning $u^*$. The local fields on the
variable are thus: $1,-1,1,2,0,0,0,0$. The satisfiable assignments are
such that $x_1=1,x_2=0,x_3=1,x_4=1$, $x_7\in\{0,1\}$, $(x_5,x_6,x_8)\in\{
(1,1,1),(1,1,0),(0,1,1),(0,0,1)\}$. One can check that the variables with nonzero
local field take the same value in all SAT assignments.
 In the WID algorithm,
the variables $1,2,3,4$ are fixed to $x_1=1,x_2=0,x_3=1,x_4=1$; the remaining
tree has only the clauses $h$ and $i$ remaining, and all the warnings are $u^*=0$.
The variable $x_7$ will be fixed arbitrarily.
If one chooses for instance $x_5=1$, the remaining graph leaves $x_8$
unconstrained but imposes $x_6=1$, as can be checked by applying to it the BP
algorithm.
}
\label{fig_WP_tree}
\end{figure}

The interpretation of the messages and the message-passing procedure
is the following. A warning $u_{a \to i}=1$ can be interpreted as a
message sent from function node $a$, telling the variable $i$ that it
should adopt the correct value in order to satisfy clause $a$. This is
decided by $a$ according to the messages which it received from all
the other variables $j$ to which it is connected: if $ \left(\sum_{b
\in V(j)\setminus a} J^b_j u_{b \to j} \right) J^a_j<0$, this means
that the tendency for site $j$ (in the absence of $a$) would be to
take a value which does not satisfy clause $a$. If all neighbors $j
\in V(a)\setminus i$ are in this situation, then $a$ sends a warning
to $i$.  An example of the use of WP is shown in
fig. \ref{fig_WP_tree}.

The warning propagation algorithm can be applied to any SAT problem.
When it converges, this dynamics defines a fixed point, which is a set
of warnings $u_{a \to i}^*$. These can be used to compute, for each
variable $i$, the ``local field'' $H_i$ and the ``contradiction
number'' $c_i$ which are two integers defined as:
\begin{equation}
H_i= -\sum_{b \in V(i)} J^b_i u_{b \to i}^*
\label{Hdef}
\end{equation}
\begin{eqnarray}
c_i&=& 1 \ \ \ \  \mbox{if}\ \  \left(\sum_{b \in V_+(i)} u_{b \to i}^*\right)
 \left(\sum_{b \in V_-(i)}u_{b \to i}^*\right) >0 \\
c_i&=& 0 \ \ \ \  \mbox{otherwise.}
\label{cdef}
\end{eqnarray}
The local field $H_i$ is an indication of the preferred
state of the variable $i$: $x_i=1$ if $H_i>0$, $x_i=0$ if
$H_i<0$. The contradiction number indicates whether the variable i
has received conflicting messages.

The interest in WP largely comes from the fact that it gives the
exact solution for tree-problems. 
This is summarized in the following simple theorem:

{\bf THEOREM 1:}

{\it Consider an instance of the SAT problem for which the
factor graph is a tree.  Then the WP algorithm converges to a unique
set of fixed point warnings $u_{a \to i}^*$, independently on the
initial warnings.  If at least one of the corresponding
contradiction numbers $c_i$ is equal to $1$, the problem is UNSAT,
otherwise it is SAT.}

{\bf Corollary:} In the case where the problem is
SAT, the local fields $H_i$ can be used to find an assignment of
the variables satisfying all the clauses, using the following 
algorithm called
 ``Warning Inspired Decimation'' or WID:
\begin{algo}
\textbf{WID} algorithm
\label{WID}

\noindent
INPUT: the factor graph of a Boolean formula in conjunctive normal
form

\noindent OUTPUT: UN-CONVERGED, or status of the formula (SAT or
UNSAT); If the formula is SAT: one assignment which satisfies all
clauses.

\linea

\begin{itemize}
\item[1.] While the number of unfixed variables is $>0$, do:
 \begin{itemize}
 \item[1.1] Run WP
 \item[1.2] If WP does not converge, return UN-CONVERGED.  Else
     compute the local fields $H_i$ and the contradiction numbers
     $c_i$, using eqs. (\ref{Hdef},\ref{cdef}).
 \item[1.3]
     If there is at least one contradiction number $c_i=1$, return UNSAT.
     Else:
     \begin{itemize}
     \item[1.3.1] If there is at least one local field $H_i \ne 0$: fix
             all variables with $H_i \ne 0$ ($H_i >0 \Rightarrow
             x_i=1$ and $H_i < 0 \Rightarrow x_i=0$ ), and clean the
             graph, which means: $\{$ remove the clauses satisfied by
             this fixing, reduce the clauses that involve the fixed
             variable with opposite literal, update the number of
             unfixed variables$\}$. GOTO label 1.  Else:
     \item[1.3.2] Choose one unfixed variable, fix it to an
           arbitrary value, clean the graph. GOTO label 1
     \end{itemize}
   \end{itemize}
 \item[2.] return the set of assignments for all the variables.
 \end{itemize}
\end{algo}

{\bf PROOF} of  theorem 1:

The convergence of message passing procedures
 on tree graphs is a well known result (see
e.g.\cite{factor_graph}). We give here an elementary proof of
convergence for the specific case of WP, and then show how the
results on $H_i$ and $c_i$ follow.

Call ${\cal E}$ the set of nodes. Define the leaves of the tree,
as the nodes of degree $1$. For any edge $(a,i)$ connecting
a function node $a$ to a variable node $i$, define
its level $r$ as follows: remove the edge $(a,i)$ and consider
the remaining subgraph containing $a$. This subgraph ${\cal T}_{a
- i}$ is a tree factor graph defining a new SAT problem.  The
level $r$ is the maximal distance between $a$ and all the leaves
in the subgraph ${\cal T}_{(a,i)}$ (the distance between two nodes 
of the graph is the number of edges of the shortest 
path connecting them). If an edge $(a,i)$ has
level $r=0$ (which means that $a$ is a leaf of the subgraph),
$u_{a \to i}(t)=1$ for all $t\ge 1$. If $(a,i)$ has level $r=1$,
then $u_{a \to i}(t)=0$ for all $t\ge 1$.  From the iteration
rule, a warning $u_{a \to i}$ at level $r$ is fully determined
from the knowledge of all the warnings $u_{b \to j}$ at
levels $\le r-2$. Therefore the warning $u_{a \to i}(t)$ at a
level $r$ is guaranteed to take a fixed value $u_{a \to i}^*$ for
$t \ge 1+ r/2$.

Let us now turn to the study of local fields and contradiction numbers.

We first prove the following {\bf lemma}:

\noindent
{\it If a warning $u_{a \to i}^*=1$, the clause $a$ is violated
in the reduced SAT problem defined by the subgraph ${\cal T}_{a \to i}$.}

\noindent 
This is obviously true if the edge has level $r=0$ or $r=1$. Supposing
that it holds for all levels $\le r-2$, one considers an edge $a -
i$ at level $r$, with $u_{a \to i}^*=1$.  From (\ref{udef}), this
means that for all variable nodes $j \in V(a)\setminus i$, the node $j$
receives at least one message from a neighboring factor node $b\in
V(j)\setminus a$ with $u_{b \to j}=1$.  The edge $b - j$ is at level $\le
r-2$, therefore the reduced problem on the graph ${\cal T}_{b - j}$
is UNSAT and therefore the clause $b$ imposes the value of the
variable $j$ to be $1$ (True) if $J_j^b=-1$, or $0$ (False)  if
$J_j^b=1$: we shall say that clause $b$ fixes the value of variable
$j$. This is true for all $j \in V(a)\setminus i$, which means that the reduced
problem on ${\cal T}_{(a,i)}$ is UNSAT, or equivalently, the clause
$a$ fixes the value of variable $i$.

Having shown that $u_{a \to i}^*=1$ implies that clause $a$ fixes the value of
variable $i$, it is clear from (\ref{cdef}) that a nonzero contradiction number
$c_i$ implies that the formula is UNSAT.

If all the $c_i$ vanish, the formula is SAT. 
One can prove this for instance by showing that the WID algorithm generates a SAT
assignment. The variables with
$H_i\ne 0$ receive some nonzero $u_{a \to i}^*=1$ and are fixed.
One then 'cleans' the graph, which means: remove the clauses
satisfied by this fixing, reduce the clauses that involve
the fixed variable with opposite literal. 
By definition, this process
has removed from the graph all the edges on which there was a nonzero warning.
So on the new graph, all the edges have $u^*=0$. Following the step 2.2 of WID,
one chooses  randomly  a variable $i$, one fixes it to an arbitrary value $x_i$,
and cleans the graph. The clauses $a$ connected to $i$ which are satisfied
by the choice $x_i$ are removed; the corresponding subgraphs are trees
where all the edges have $u^*=0$. A clause $a$ connected to $i$ which
 are not satisfied by the choice $x_i$ may send some $u^*=1$ messages
(this happens if such a clause had degree $2$ before fixing variable $i$).
However, running WP on the corresponding subgraph ${\cal T}_{(a,i)}$, the set
of warnings can not have a contradiction: A variable
$j$ in this subgraph can receive at most one $u^*=1$ warning, coming
from the unique path which connects $j$ to $a$. Therefore  $c_j=0$: 
one iteration of WID has generated
a strictly smaller graph with no contradiction. By induction,
it thus finds a SAT assignment.$\square$

One should notice that the variables which are fixed
at the first iteration of WID (those with non-zero $H_i$) 
are constrained to take the same value in all
satisfiable assignments.

\subsection{Belief propagation}
While the WP algorithm is well adapted to finding a SAT assignment,
the more complicated belief propagation (BP) algorithm  is able to
compute, for satisfiable problems with a tree factor graph, the 
total number of SAT assignments, and the
fraction of SAT assignments where a given variable $x_i$ is true.

We consider a satisfiable instance, and  the probability space built by 
all SAT assignments taken with equal probability.
Calling $a$ one of the clauses in which $x_i$ appears,
the basic ingredients of BP are the messages: 
\begin{itemize}
\item $\mu_{a\to i}(x_i) \in [0,1]$, interpreted as 
the probability that clause $a$ is satisfied,
given the value  of the variable $x_i\in\{0,1\}$.
\item $\mu_{i \to a}(x_i) \in [0,1]$, interpreted as the probability that the variable takes value $x_i$,
when clause $a$ is absent (this is again a typical 'cavity'
definition).
Notice that $\sum_{x_i \in \{0,1\}}\mu_{i \to a}(x_i)=1$,
while there is no such normalization for $\mu_{a\to i}(x_i)$.
\end{itemize}
The BP equations are:
\begin{equation}
\mu_{i \to a}(x_i)=C_{i \to a} \prod_{b \in V(i) \setminus a} \mu_{b\to i} (x_i) \ ,
\label{bp1}
\end{equation}
\begin{equation}
\mu_{a \to i} (x_i)= 
\sum_{ \{ x_j (j\ne i) \} } f_a(X) \prod_{j \in V(a)\setminus i} \mu_{j \to a} (x_j) \ ,
\label{bp2}
\end{equation}
where $C_{i \to a}$ is a normalization constant ensuring that $\mu_{i \to a}$
is a probability, the sum over $ \{x_j  (j\ne i) \}$ means a sum over 
all values of the variables $x_j \in \{0,1\}$, for all $j$ different from $i$,
and $f_a(X)$ is a characteristic function taking value $1$ if
the configuration $X=\{x_i\}$ satisfies clause $a$, taking value $0$ 
otherwise.

 It is convenient to  parameterize $\mu_{i \to a}(x_i)$ by introducing 
the number $\gamma_{i \to a} \in [0,1]$ which is the probability that 
 the variable $x_i$ is in the state which violates clause $a$, in a problem
where clause $a$ would be absent (writing for instance $ \mu_{i \to a}(x_i)=
\gamma_{i \to a} \delta(x_i,0)+ (1-\gamma_{i \to a}) \delta(x_i,1)$
in the case where $J_i^a=-1$). 

Let us denote by
\begin{equation}
\delta_{a \to i}\equiv \prod_{j \in V(a)\setminus i} \gamma_{j \to a}
\end{equation}
the probability that all variables in clause $a$, except variable $i$,
are in the state which violates the clause. 

The BP algorithm amounts to an iterative update of the 
messages $\delta_{a \to i}$
according to the rule:

\begin{algo}
\noindent\textbf{BP} algorithm:
\label{BP}
\noindent INPUT: the factor graph of a Boolean formula in conjunctive normal
form; a maximal number of iterations $t_{max}$; a requested precision
$\epsilon$.

\noindent OUTPUT: UN-CONVERGED if BP has not converged after $t_{max}$
sweeps. If it has converged: the set of all 
messages $\delta_{a \to i}^*$.

\linea
\begin{itemize}
\item[0.] At time $t=0$: For every edge $a \to i$ of the factor graph,
randomly initialize the messages $\delta_{a \to i}(t=0) \in [ 0,1]$
\item[1.] For  $t= 1$ to $t=t_{max}$:
\begin{itemize}
\item[1.1] sweep the set of edges in a random order, and update
 sequentially the warnings on all the edges of the graph,
 generating the values $\delta_{a \to i}(t)$, using subroutine
 BP-UPDATE.
\item[1.2] If $|\delta_{a \to i}(t)-\delta_{a \to i}(t-1)|<\epsilon $
on all the edges, the iteration has converged and generated $\delta_{a \to
i}^*=\delta_{a \to i}(t)$: go to 2.
\end{itemize}
\item[2.] If $t=t_{max}$ return UN-CONVERGED. If $t<t_{max}$ return
the set of fixed point warnings $\delta{a \to i}^*=\delta_{a \to i}(t)$
\end{itemize}

\end{algo}

\begin{algo}
\noindent Subroutine BP-UPDATE$(\delta_{a \to i})$

\noindent INPUT: Set of all messages arriving onto each variable node $j \in
V(a) \setminus i$

\noindent OUTPUT: new value for the message $\delta_{a \to i}$.

\linea

\begin{itemize}
\item[1] For every $j \in V(a) \setminus i$, compute the cavity field
\begin{equation}
\gamma_{j \to a}=\frac{P^u_{j\to a}}{P^u_{j\to a}+ P^s_{j\to a}}
\end{equation}
where
\begin{eqnarray} \nonumber
P^u_{j\to a}&=&\prod_{b \in V_a^s(j)}(1-\delta_{b \to j}) \ ,\\
P^s_{j\to a}&=&\prod_{b \in V_a^u(j)}(1-\delta_{b \to j}) \ .
\label{bp2prime}
\end{eqnarray}
If an ensemble is empty, for instance $ V_a^s(j)=\emptyset$, the
corresponding $P^u_{j\to a}$ takes value $1$ by definition.

\item[2] Using these numbers $\gamma_{j \to a}$, compute the new
message: $\delta_{a \to i}\equiv \prod_{j \in V(a)\setminus i}
\gamma_{j \to a}$.  If a factor node $a$ is a leaf (unit clause) with
a single neighbor $i$, the corresponding $\delta_{a \to i}$ takes
value $1$ by definition.
\end{itemize}

\end{algo}

\begin{figure}
\centering
\includegraphics[width=8.cm]{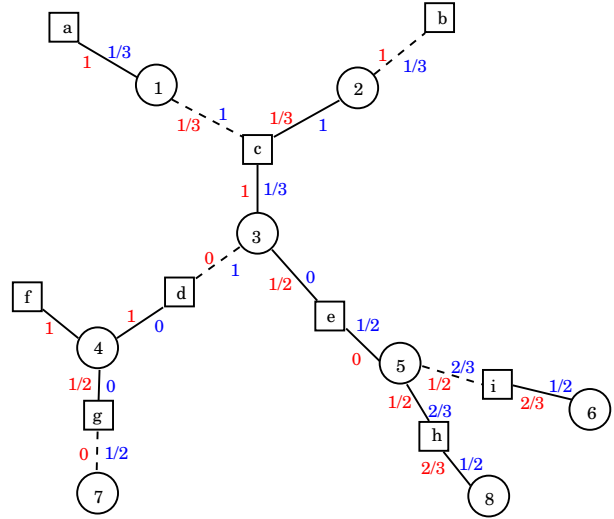}
\caption{ An example of result obtained by the BP algorithm on the
 tree-problem with $N=8$ variables and $M=9$ clauses studied in
 fig.\ref{fig_WP_tree}.  On each edge of the graph connecting a
 function node like $c$ to a variable node like $2$, appears the value
 of $\gamma_{2 \to c}$ (in blue, on the right hand side of the edge)
 and of $\delta_{c \to 2}$ (in red, on the left hand side of the
 edge). Comparing to the WP result of fig.\ref{fig_WP_tree}, one sees
 that all the messages $\delta_{a \to i}=1$, corresponding to strict
 warnings, are the same, while the messages $\delta_{a \to i}<1$ are
 interpreted in WP as ``no warning'' (i.e. a null message).  Using
 (\ref{mucalc}), the probability $\mu_i$ that each variable $x_i=1$
 are found equal to: $1,0,1,1,1/2,3/4,3/4,1/2$. These are the exact
 results as can be checked by considering all satisfiable assignments
 as in fig.\ref{fig_WP_tree}.  }
\label{fig_BP_tree}
\end{figure}
As WP, the BP algorithm is exact on trees (see for instance
\cite{factor_graph}).  In fact it gives a more accurate results than
WP since it allows to compute the exact probabilities $\mu_i$ (while
WP identifies the variables which are fully constrained, and gives a
zero local field on the other variables). A working example of BP is
shown in fig.\ref{fig_BP_tree}. In this example and more in general
for trees, BP also provides the exact number ${\cal N}$ of SAT
assignments, as given by the following theorem:

{\bf THEOREM 2:} 
{\it Consider an instance of the SAT problem for which the
factor graph is a tree, and there exist some SAT assignments. Then:

a) The BP
algorithm converges to a unique set of fixed point messages $\delta^*_{a \to
i}$. 

b)The probability $\mu_i$ that the variable $x_i=1$ is given by:}
\begin{equation}
\mu_i=\frac{\prod_{a \in V_-(i)}(1-\delta^*_{a \to i})}{
\prod_{a \in V_-(i)}(1-\delta^*_{a \to i})+
\prod_{a \in V_+(i)}(1-\delta^*_{a \to i})} \ .
\label{mucalc}
\end{equation}
{\it c) The number of SAT assignments is ${\cal N}=\exp(S)$, where
 the entropy $S$ is given by:}
\begin{eqnarray}\nonumber
S&=&\sum_{a \in A} \ln
\left[
	\prod_{i \in V(a)} 
	\left(
		\prod_{b\in V_a^s(i)} (1-\delta_{b \to i}^*)+
		\prod_{b\in V_a^u(i)} (1-\delta_{b \to i}^*)
	\right)
- \prod_{i \in V(a)} 
	\left(
		\prod_{b\in V_a^u(i)} (1-\delta_{b \to i}^*)
	\right)
\right]\\
&&+\sum_{i \in X} (1-n_i) \ln
\left[
	\prod_{b\in V_+(i)} (1-\delta_{b \to i}^*)+ 
	\prod_{b\in V_-(i)} (1-\delta_{b \to i}^*)
\right]
\label{entrop_BP}
\end{eqnarray}

{\bf PROOF:} 

The proof of convergence is simple, using the same strategy as
 the proof in sect.\ref{sect_WP}: messages at level 0 and 1 are fixed
automatically, and a  message at level $r$ is fixed by the values of messages
at lower levels.

The probability $\mu_i $ is computed from the same procedure as
the one giving the BP equations (\ref{bp1},\ref{bp2}), with the difference 
that one takes into account all neighbors of the site $i$.

The slightly more involved result is the one concerning the entropy.
We use the probability measure $P(X)$ on the space of all assignments
which has uniform probability for all SAT assignments and zero probability
for all the assignments which violate at least one clause:
\begin{equation}
P(X)=\frac{1}{\cal N} \prod_{a \in A} f_a(X) \ .
\end{equation}
From $P$ one can define  the following marginals: 
\begin{itemize}
\item the 'site marginal' $p_i(x_i)$ is the
probability that variable $i$ takes value $x_i \in \{ 0,1\}$
\item the 'clause marginal' $p_a(X_a)$
is the probability that the set of variables $x_i, i \in V(a)$, takes
a given value, denoted by $X_a$ (among the $2^{n_a}$ possible values).
\end{itemize}
For a tree factor graph, one easily shows by induction on the size of the graph
that  the full
probability can be expressed in terms of the site and clause marginals as:
\begin{equation}
P(X)=\prod_{a\in A}  p_a(X_a) \prod_{i \in X} p_i(x_i)^{1-n_i} \ .
\end{equation}
The entropy $S= \ln({\cal N})$ is then obtained as
\begin{equation}
S=-\sum_X P(X)\ln P(X)=-\sum_a \sum_{X_a} p_a(X_a) \ln[ p_a(X_a) ]-
\sum_i (1-n_i) \sum_{x_i} p_i(x_i) \ln[ p_i(x_i) ]
\end{equation}
Let us now derive the expression of this quantity in terms of the
messages used in BP. 
One has:
\begin{equation}
 p_i(x_i)= c_i \prod_{b \in V(i)} \mu_{b \to i}(x_i)
\label{pi}
\end{equation}
and
\begin{equation}
 p_a(X_a)= c_a  f_a(X_a)\prod_{i \in V(a)} \mu_{i \to a}(x_i) \ ,
\label{pa}
\end{equation}
where $c_i$ and $c_a$ are two normalization constants.
From (\ref{pi}) one gets after some reshuffling:
\begin{equation}
\sum_i (n_i-1) \sum_{x_i} p_i(x_i) \ln p_i(x_i)=
\sum_i (n_i-1)\ln c_i+\sum_a \sum_{i \in V(a)} \sum_{X_a}  p_a(X_a)
\ln  \left[\prod_{b \in V(i)\setminus a
} \mu_{b \to i}(x_i)\right] \ ;
\end{equation}
Using the BP equation (\ref{bp1}), this gives:
\begin{equation}
\sum_i (n_i-1) \sum_{x_i} p_i(x_i) \ln p_i(x_i)=
\sum_i (n_i-1)\ln c_i+\sum_a  \sum_{X_a}  p_a(X_a)
\ln \left[\prod_{i \in V(a)} \mu_{i \to a}(x_i)  f_a(X_a)\right] 
-\sum_a \sum_{i \in V(a)} \ln C_{i \to a}
\ ,
\end{equation}
where the term $ f_a(X_a)$ inside the logarithm has been added, taking into
account the fact that, as $ f_a(X_a) \in \{0,1\}$, one always has 
$p_a(X_a) \ln f_a(X_a) =0$.
Therefore:
\begin{equation}
S=-\sum_a \ln c_a+ \sum_i(n_i-1) \ln c_i -
 \sum_a \sum_{i \in V(a)} \ln C_{i \to a}
\label{s_interm}
\end{equation}
In the notations of (\ref{bp2prime}), one has
\begin{equation}
c_a=\frac{1}{1-\prod_{i \in V(a)} \gamma_{i \to a}}=
 \frac{1}{1-\prod_{i \in V(a)} P^u_{i \to a}/(P^u_{i \to a}+P^s_{i \to a})} \ ,
\end{equation}
\begin{equation}
C_{i \to a}=\frac{1}{P^u_{i \to a}+P^s_{i \to a}} \ ,
\end{equation}
and
\begin{equation}
c_i=\frac{1}{\prod_{b\in V_+(i)}(1-\delta_{b \to i})+ 
\prod_{b\in V_-(i)}(1-\delta_{b \to i})} \ .
\end{equation}
Substitution into (\ref{s_interm}) gives the expression (\ref{entrop_BP}) 
for the entropy. $\square$

\section{Survey Propagation}
\label{wptosp}
\subsection{The algorithm}
The WP and BP algorithms have been shown to work for satisfiability problems
where the factor graph is a tree. In more general cases where the factor graph
has loops, they can be tried as heuristics, but there is no guarantee of
convergence. In this section we present a new message passing algorithm,
survey propagation (SP), which is also a heuristic, without any guarantee of
convergence. It reduces to WP for tree-problems,
but it turns out to be more efficient than WP or BP in experimental
studies of random satisfiability problems. SP has been discovered using
concepts developed in statistical physics under the name of 'cavity method'.
Here we shall first present SP, then give some experimental results, and in
the end expose the qualitative physical reasoning behind it.

A  message of SP, called a  survey,
passed from one function node $a$ to a
variable $i$ (connected by an edge) is a real
number $\eta_{a \to i} \in [0,1]$. The SP algorithm
uses exactly the same main procedure as  BP (see \ref{BP}),
but instead of calling the BP-UPDATE, it
uses a different update rule, SP-UPDATE, defined as:

\begin{algo}
Subroutine SP-UPDATE$(\eta_{a \to i})$

\noindent
INPUT: Set of all messages arriving onto each variable node $j \in
V(a) \setminus i$

\noindent
OUTPUT: new value for the message $\eta_{a \to i}$.

\linea

\begin{itemize}
\item[1] For every $j \in V(a) \setminus i$, compute the three
numbers:
\begin{eqnarray}
\nonumber \Pi^u_{j\to a}&=& \left[1-\prod_{ b\in
 V^u_a(j)}\left(1-\eta_{b\to j}\right)\right] \prod_{ b\in V^s_a (j)
 }\left(1-\eta_{b\to j}\right) \\ \nonumber \Pi^s_{j\to a}&=&
 \left[1-\prod_{ b\in V^s_a(j)}\left(1-\eta_{b\to j}\right)\right]
 \prod_{ b\in V^u_a (j) }\left(1-\eta_{b\to j}\right)\\ \Pi^0_{j\to
 a}&=& \prod_{ b\in V(j)\setminus a }\left(1-\eta_{b\to j}\right)
\label{eta2}
\end{eqnarray}
 if a set like $ V^s_a(j)$ is empty, the corresponding product takes
value $1$ by definition.

\item[2] Using these numbers , compute and return the new survey:
\begin{equation}
 \eta_{a \to i}=\prod_{j \in V(a)\setminus i} \left[ \frac{\Pi^u_{j\to
a}} {\Pi^u_{j\to a}+\Pi^s_{j\to a}+\Pi^0_{j\to a}} \right] \ .
\label{eta1}
\end{equation}
If $V(a) \setminus i$ is empty, then $\eta_{a \to i}=1$.

\end{itemize}
\end{algo}

Qualitatively, the statistical physics interpretation 
of the survey $\eta_{a \to i}$ is a probability that a warning 
is sent from $a$ to $i$ (see section \ref{heuristic} for
details). Therefore, whenever the SP algorithm converges to
a fixed-point set of messages $\eta_{a \to i}^*$, one can use it 
in a decimation procedure in order to find a satisfiable assignment,
if such an assignment exists. This procedure, called the survey inspired 
decimation (SID), is a generalization of the WID algorithm \ref{WID}, 
defined by:

\linea

\begin{sf}

\noindent{\bf SID} algorithm
\label{SID}

\noindent INPUT: The factor graph of a Boolean formula in conjunctive normal
form.  A maximal number of iterations $t_{max}$ and a precision
$\epsilon$ used in SP

\noindent OUTPUT: One assignment which satisfies all clauses, or 'SP
UNCONVERGED', or 'probably UNSAT'
\linea

\begin{itemize}
\item[0.] Random initial condition for the surveys

\item[1.] Run SP.  {\bf If SP does not converge}, return 'SP
UNCONVERGED' and stop (or restart, i.e.{\bf go to 0.}).  {\bf If SP
converges}, use the fixed-point surveys $\eta^*_{a \to i}$ in order to:

\item[2.] Decimate:
\begin{itemize}

\item[2.1] {\bf If non-trivial surveys ($\{ \eta \neq 0\}$) are
found}, then: 
\begin{itemize} 
 \item[(a)]Evaluate, for each variable node $i$, 
 the three 'biases' $\{W_i^{(+)}, W_i^{(-)}, W_i^{(0)}\}$defined by:
\begin{eqnarray}
W_i^{(+)} &=& \frac{\hat\Pi_i^+}{\hat\Pi^+_{i}+\hat\Pi^-_{i}+\hat\Pi^0_{i}} \\
W_i^{(-)} &=& \frac{\hat\Pi_i^-}{\hat\Pi^+_{i}+\hat\Pi^-_{i}+\hat\Pi^0_{i}} \\
W_i^{(0)} &=& 1-W_i^{(+)}-W_i^{(-)}
\label{wdef}
\end{eqnarray}
where $\hat \Pi_i^+,\hat\Pi_i^-,\hat\Pi_i^0$ are defined by
\begin{eqnarray}
 \hat \Pi^+_{i}&=&
\left[1-\prod_{ a\in V_+(i)}\left(1-\eta^*_{a\to i}\right)\right]
\prod_{ a\in V_- (i) }\left(1-\eta^*_{a\to i}\right) \nonumber \\
\hat \Pi^-_{i}&=&
\left[1-\prod_{ a\in V_-(i)}\left(1-\eta^*_{a\to i}\right)\right]
\prod_{ a\in V_+ (i) }\left(1-\eta^*_{a\to i}\right) \nonumber \\
\hat \Pi^0_{i}&=&
\prod_{ a\in V(i) }\left(1-\eta^*_{a\to i}\right)
\label{hatpidef}
\end{eqnarray}

 \item[(b)]\underline{fix} the variable with the largest $\vert
W_i^{(+)}- W_i^{(-)}\vert$ to the value $x_i=1$ if $W_i^{(+)}>
W_i^{(-)}$, to the value $x_i=0$ if $W_i^{(+)}< W_i^{(-)}$.  Clean the
graph, which means: $\{$ remove the clauses satisfied by this fixing,
reduce the clauses that involve the fixed variable with opposite
literal, update the number of unfixed variables$\}$.
\end{itemize}
\item[2.2] {\bf If all surveys are trivial ($\{ \eta=0\}$) }, then
output the simplified sub-formula and run on it a local search process
(e.g. walksat).

\end{itemize}

\item[4.] If the problem is solved completely by unit clause
propagation, then output ``SAT'' and stop.  If no contradiction is
found then continue the decimation process on the smaller problem
({\bf go to 1.})  else (if a contradiction is reached) stop.
\end{itemize}
\end{sf}
\linea
\bigskip

There exist several variants of this algorithm. In the code which is
available at \cite{web}, for performance reasons we update
simultaneously all $\eta$ belonging to the same clause. The clauses to
be updated are chosen in a random permutation order at each iteration
step. The algorithm can also be randomized by fixing, instead of the
most biased variables, one variable randomly chosen in the set of the
x percent variables with the largest bias.  This strategy allows to
use some restart in the case where the algorithm has not found a
solution. A fastest decimation can also be obtained by fixing in the
step 2.1(b), instead of one variable, a fraction $f$ of the $N_t$
variables which have not yet been fixed (going back to $1$ variable
when $f N_t<1$).

\subsection{Experimental study of the SP algorithm}

\label{sect_experiment}

In order to get some concrete information on 
the behaviour of SP for large
but finite $N$, we have experimented SP and SID on single instances of
the random 3-SAT problem with many variables, up to $N\sim 10^7$. In
this section we summarize these (single machine) experiments and their
results.

Instances of the 3-SAT problem were generated with the pseudo random
number generator "Algorithm B" on p.32 of Knuth~\cite{Knuth}.  However
we found that results are stable with respect to changes in the random
number generators. Formulas are generated by choosing k-tuples of
variable indices at random (with no repetitions) and by negating
variables with probability $0.5$.

We first discuss the behaviour of the SP algorithm itself. We have
used a precision parameter $\epsilon=10^{-3}$ (smaller values don't
seem to increase performance significantly).  Depending on the range
of $\alpha$, we have found the following behaviours, for large enough
$N$:

\begin{itemize}
\item For $\alpha <\alpha_d\sim 3.9$, SP converges towards the set of
trivial messages $\eta_{a \to i}=0$, for all $a-i$ edges.  All
variables are under-constrained.

\item For $3.9 < \alpha < 4.3$, SP converges to a unique
fixed-point set of  non-trivial messages, independently from the initial
conditions, where a large fraction of the messages $\eta_{a \to i}$ 
are in $]0,1[$.  

\end{itemize}

Notice that, for `small' values of $N$, around $N=1000$, one often
finds some instances in which SP does not converge. But the
probability of convergence, at a given $\alpha<4.3$, increases with
$N$. This is exemplified by the following quantitative measure of the
performance of the SID algorithm (which uses SP).  We have solved
several instances of the random 3-SAT problem, for various values of
$\alpha$ and $N$, using the SID algorithm in which we fix at each step
the fraction $f N_t$ of variables with largest $\left|
W_i^{(+)}-W_i^{(-)} \right|$. Table \ref{tab} gives in each case the
fraction of samples which are solved by SID, in a single run of
decimation (without any restart). The algorithm fails when, either SP
does not converge, or the simplified sub-formula found by SID is not
solved by walksat.  The performance of SID improves when $N$ increases
and when $f$ decreases. Notice that for $N=10^5$ we solve all the $50$
randomly generated instances at $\alpha=4.24$. For larger values of
$\alpha$ the algorithm often fails.  Notice that in such cases it does
not give any information on whether the instance is UNSAT. Some
failures may be due to UNSAT instances, others are just real failures
of the SID for SAT instances.  Few experiments on even larger
instances ($N=10^6,10^7$) have also been succesfully run (the limiting
factor being the available computer memory needed to store formulas).

\begin{figure}[Table]
\label{tab}
\centering
\input{table.tex}
\caption{ Results obtained by solving with a single decimation run of
the SID algorithm 50 random instances of 3-SAT with
$N=25000,50000,100000$ and $\alpha=4.21,4.22,4.23,4.24$.  SID was used
by fixing variables in blocks $f N_t$, where $N_t$ is the number of
unfixed variables at time $t$, and various runs used different values
of $f$ taken in the geometric progression $f=4 \%,2 \%,1 \%,.5 \%,.25
\%,.125 \%$, stopping if the formula was solved. For each value of
$(N, \alpha, f)$, we give the fraction of the 50 instances which were
solved (i.e. for which the algorithm found a SAT assignement).  The
maximal number of iteration was taken equal to $10^3$ and the
precision for convergence was taken equal to $10^{-3}$.  The last row
shows the number of complete iterations of SP averaged over the
successful runs.}
\end{figure}

As shown by the data on the average total number of SP iterations
along the successful solution process in table \ref{tab}, the
convergence time of the SP algorithm basically does not grow with $N$
(a growth like $\log N$, which could be expected from the geometrical
properties of the factor graph, is not excluded).  Therefore the
process of computing all the SP messages $\eta^*_{a \to i}$ takes
$\Theta(N)$, or maybe $\Theta(N\ln N)$, operations. If SID fixes at
each step only one variable, it will thus converge in $\Theta(N^2 \log
N)$ operations (the time taken by walksat to solve the simplified
sub-formula seems to grow more slowly).  When we fix a fraction of
variables at a time, we get a further reduction of the cost to $O(N
(\ln N)^2)$ (the second $\ln$ comes from sorting the biases).

A very basic yet complete version of the code which is intended to
serve only for the study on random 3-SAT instances is available at the
web site \cite{web}.  Generalization of the algorithm to other problems 
require some changes which are not implemented in the distributed code.

\section{Heuristic arguments}
\label{heuristic}
Survey propagation has been invented using powerful concepts and methods
developed in the statistical physics of disordered systems, notably
the cavity method for diluted problems \cite{Bethe_cav}. In this section
we want to give some short background on these methods, in order
to help the reader understand where SP comes from, and maybe develop
similar algorithms in other contexts. Unfortunately so far there is
no rigorous derivation of the cavity method, so this whole section
only contains heuristic arguments.

\subsection{The physical picture underlying the SP construction: clustering 
of configurations}
Let us start with a discussion of the validity of BP. As we saw in
sect.\ref{sect_WP}, BP aims at computing the marginal probability
distribution of a variable $x_i$, within the probability space built
by all SAT assignments, each being given equal probability. The
message $\mu_{a \to i}(x_i)$ used in BP can be computed exactly if one
knows the joint probability distribution $ P^{(a)}(X) $ of the
variables in $X=\{ x_j, \; j \in V(a)\setminus i\}$, in the graph
where clause $a$ is absent. Using the same notations as in
(\ref{bp2}), one has:
\begin{equation}
\mu_{a \to i} (x_i)= 
\sum_{ \{ x_j (j\ne i) \} } f_a(X) P^{(a)} (X)
\label{bpexact}
\end{equation}
Comparing this eq. (\ref{bpexact}) to the eqs. (\ref{bp1},\ref{bp2})
of BP, one sees that BP uses an approximation, namely the fact that
the joint probability $ P^{(a)} (X)$ factorizes: $ P^{(a)} (X) \simeq
\prod_{j \in V(a)\setminus i} \mu_{j \to a} (x_j)$.  This amounts to
assuming that the variables $x_j$, for $ j \in V(a)\setminus i$, are
uncorrelated in the absence of clause $a$. This assumption is
obviously correct when the factor graph is a tree, which confirms the
validity of BP in that case. In a general problem, for such an
assumption to hold, we need two conditions to be fulfilled:
\begin{itemize}
\item
the variables $ x_j, j \in V(a)\setminus i$ should be far from each
other, in the factor graph where clause $a$ is absent.
\item 
there should be a single phase in the problem, or the probability
measure should be reduced to one single pure phase.
\end{itemize}
The first condition is easily understood, and one can see that it is
generically fulfilled when one considers a random satisfiability
problem. In random K-sat with $M=\alpha N$ clauses, the factor graph
is a random bipartite graph, where function nodes have degree $K$ and
variable nodes have fluctuating degrees, with a distribution which
becomes, in the large $N$ limit, a Poisson law of mean $K
\alpha$. Locally such a graph is tree like.  Taking a clause $a$ at
random, the minimal distance between two of the variables $ x_j, j \in
V(a)\setminus i$ is generically of order $\log N$.

The second condition is more subtle. In general a pure phase is
defined in statistical physics as an extremal Gibbs measure
\cite{georgii}; however the standard construction of Gibbs measures
deals with infinite systems. Here we need to work with $N$ variables
where $N \gg 1$ but is finite, and the corresponding construction has
not been worked out yet. For the random satisfiability problem, a
heuristic description of a pure phase is a cluster of SAT assignments,
defined as follows. Consider all SAT assignments. Define the distance
between two assignments $\{ x_i \}$ and $\{ y_i \}$ as $\sum_i
(x_i-y_i)^2$. If the distance between two SAT assignment is smaller
than a number $q$, they are said to belong to the same
$q$-cluster. This allows to partition the set of SAT assignments into
$q$-clusters. One is interested in the 'clusters' obtained in large
$N$ --  large $q$ limit of the $q$-clusters, where the large $N$ limit is
taken first. Heuristic statistical physics arguments indicate that, in
the random K-satisfiability problem, for $\alpha<\alpha_{clust}(K)$,
there should exist one single such cluster of SAT assignments: this
means that one can move inside the space of SAT assignment, from any
assignment to another one, by a succession of moves involving each a
number of flips of variables which is $<<N$. In such a case the BP
factorization approximation is expected to be correct. On the other
hand, for $\alpha>\alpha_{clust}(K)$ the space of SAT assignment
separates into many distant clusters, and the BP factorization does
not hold globally, but it would hold if one could restrict the
probability space to one given cluster $\alpha$.  Within such a
restricted space, BP would converge to a set of messages $\mu_{a \to
i}^\alpha(x_i)$ which depends on the cluster $\alpha$.

In this situation, the cavity method uses a statistical approach.  It
considers all the clusters of SAT assignments, and attributes to each
cluster a probability proportional to the number of configuration that
it contains. Then one introduces, on each edge $a-i$, a survey which
gives the probability, when a cluster $\alpha$ is chosen randomly with
this probability, that the message $\mu_{a \to i}^\alpha(x)$ is equal
to a certain function $P(x)$.

This object is a probability of a probability and it is thus difficult
to use in practical algorithms. For this reason, SP departs from the
usual cavity method and uses a simpler
object which is a survey of warnings, interpreted as follows: Consider
one cluster $\alpha$ and an edge $a-i$ of the factor graph.  If, in
every SAT assignments of the cluster $\alpha$, all the variables $x_j,
j \in V(a)\setminus i$ don't satisfy clause $a$, then a warning $u_{a
\to i}^\alpha=1$ is passed along the edge from $a$ to $i$.  The SP
message along this edge is the survey of these warnings, when one
picks up a cluster $\alpha$ at random: $\eta_{a \to i}= \sum_\alpha
u_{a \to i}^\alpha/(\sum_\alpha 1)$. So basically the SP message gives
the probability that there is a warning sent from $a$ to $i$ in a
randomly chosen cluster.  With respect to the full-fledged cavity
method, this is a much simplified object, which focuses onto the
variables which are constrained.

The experimental results on random 3-satisfiability discussed in
sect. \ref{sect_experiment} confirm the theoretical analysis of
\cite{MEPAZE,MZ_pre}
 which indicate that all $\eta_{a \to i}$ vanish for
$\alpha<\alpha_d \simeq 3.91$. This can be interpreted as the fact
that, in this range of $\alpha$, there are no 'constrained clusters'
(meaning clusters in which some of the variables are constrained). For
$\alpha_d<\alpha<\alpha_c=4.267$ the theory predicts the existence of
non-trivial messages, meaning that there exist constrained clusters.
This is the region where SP and SID are able to outperform existing
algorithms.  One should notice that the SID does not seem to converge
up to the conjectured threshold $\alpha_c=4.267$: although it is very
small, there remains a region in the 'Hard SAT' phase, close to
$\alpha_c$ where the simple version of SP/SID does not give the result.
Recent work shows that this gap can be reduced by using some
backtracking strategy in SID\cite{Parisi_backtrack}. Whether it will be
possible to close the gap with generalized versions of SP/SID, while
keeping a typically polynomial running time, is an interesting open
issue.

\subsection{The ``don't-care'' state}
In a given cluster $\alpha$, a variable $x_i$ can be thought of being
in three possible states: either it is constrained equal to $0$ (this
means that $x_i=0$ in all SAT assignments of the cluster $\alpha$), or
it is constrained equal to $1$, or it is not constrained. In this last
situation we attribute it the value $*$. Therefore we can describe a
cluster by the values of the $N$ generalized variables $x_i^\alpha \in
\{ 0,1,*\}$, where $*$ will be denoted as the don't-care state.  Such
a description associates to each cluster a single point in $\{
0,1,*\}^N$.  It discards a lot of information: $x_i^\alpha =*$ has
lost all the information on the fraction of assignments in the cluster
$\alpha$ where $x_i=0$. But it gives a simplified description of the
cluster and it focuses onto the constrained variables.

It is interesting to notice that the SP equations can be interpreted
as BP equations in the presence of this extra don't-care state. This
can be seen as follows.  Borrowing the notations of the BP equations
(\ref{bp1}, \ref{bp2}) we denote by $\gamma_{i \to a} \in [0,1]$ the
probability that the variable $x_i$ is in the state which violates
clause $a$, in a problem where clause $a$ would be absent, and by
\begin{equation}
\delta_{a \to i}= \prod_{j \in V(a)\setminus i} \gamma_{j \to a}
\label{spn1}
\end{equation}
 the probability that all variables in clause $a$, except variable
$i$, are in the state which violates the clause.

Let us compute $\mu_{i \to a}(x_i)$. This depends on the messages sent
from the nodes $b \in \{ V(i)\setminus a\} $ to variable $i$. The
various possibilities for these messages are:
\begin{itemize}
\item
No warning arriving from  $b \in V_a^s(i)$, and no warning
arriving from $b \in V_a^u(i)$. This happens with a probability
\begin{equation}
\Pi^0_{i \to a}=\prod_{ b\in V(i)\setminus a }\left(1-\delta_{b\to i}\right)
\label{spn2}
\end{equation}
\item
No warning arriving from  $b \in V_a^s(i)$, and at least one warning
arriving from $b \in V_a^u(i)$. This happens with a probability
\begin{equation}
\Pi^u_{i \to a}=
\left[1-\prod_{ b\in V^u_a(i)}\left(1-\delta_{b\to i}\right)\right]
\prod_{ b\in V^s_a (i) }\left(1-\delta_{b\to i}\right)
\label{spn3}
\end{equation}
\item
No warning arriving from  $b \in V_a^u(i)$, and at least one warning
arriving from $b \in V_a^s(i)$. This happens with a probability
\begin{equation}
\Pi^s_{i \to a}=
\left[1-\prod_{ b\in V^s_a(i)}\left(1-\delta_{b\to i}\right)\right]
\prod_{ b\in V^u_a (i) }\left(1-\delta_{b\to i}\right)
\label{spn4}
\end{equation}
\item
At least one warning arriving from  $b \in V_a^u(i)$, and at least one warning
arriving from $b \in V_a^s(i)$. This happens with a probability
\begin{equation}\Pi^c_{i \to a}=
\left[1-\prod_{ b\in V^s_a(i)}\left(1-\delta_{b\to i}\right)\right]
\left[1-\prod_{ b\in V^u_a(i)}\left(1-\delta_{b\to i}\right)\right]
\label{spn5}
\end{equation}
\end{itemize}

As we work only with SAT configurations, the contradictory messages
must be excluded. Therefore, the probability $\gamma_{i \to a} \in
[0,1]$ that the variable $x_i$ is in the state which violates clause
$a$, given that there is no contradiction, is:
\begin{equation}
\gamma_{i \to a}=\Pi^u_{i \to a}/(\Pi^u_{i \to a}+\Pi^s_{i \to
a}+\Pi^0_{i \to a})
\label{spn6}
\end{equation}
The above equations in the enlarged space including the null message,
given in (\ref{spn1}-\ref{spn6}) are identical to the SP
equations(\ref{eta1},\ref{eta2}), with the identification $\eta_{a \to
i} =\delta_{a \to i}$.

\subsection{Complexity}

The above interpretation of SP using the don't-care state suggest a
method to estimate the complexity, defined as the normalized logarithm
of the number of constrained clusters of SAT assignments.  As each
constrained cluster of SAT assignments is associated with a point in
$\{ 0,1,*\}^N$, the complexity should be given by the corresponding
entropy, which can be estimated with the usual BP formula
(\ref{entrop_BP}) in the presence of the don't-care state (see
ref. \cite{BZ_jstat} for a rigorous derivation).  The result,
originally derived in \cite{MZ_pre} from a direct statistical physics
analysis without the use of the don't-care state, is as follows.

The total complexity $\Sigma$ can be decomposed into contributions
associated with every function node and with every variable, and
reads:
\begin{equation}
\Sigma= \sum_{a=1}^M \Sigma_a -\sum_{i=1}^N (n_i-1)\Sigma_i
\label{sigma_def}
\end{equation}
where
\begin{equation}
\Sigma_a=+\log \left[
        \prod_{j \in V(a)}
    \left(\Pi^u_{j\to a}+\Pi^s_{j\to a}+\Pi^0_{j\to a}\right)-
    \prod_{j \in V(a)}
  \Pi^u_{j\to a}
\right]
\label{sigma_bond}
\end{equation}
\begin{equation}
\Sigma_i=+\log\left[ \hat \Pi^+_{i}+ \hat \Pi^-_{i}+ \hat \Pi^0_{i} \right]
\label{sigma_site}
\end{equation}

\begin{figure}
\centering
\includegraphics[width=8.cm]{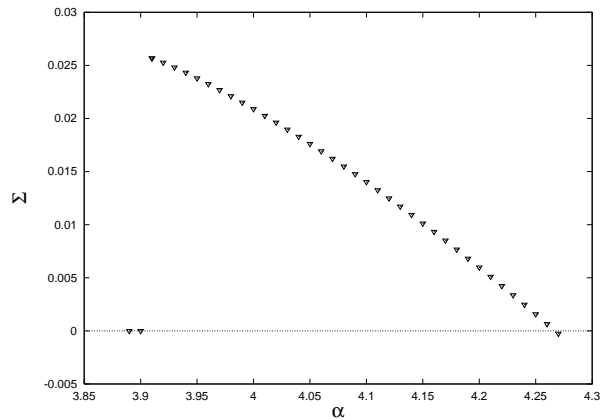}
\caption{Complexity per variable ($\Sigma/N$) of the satisfying
clusters for a given sample of size $N=10^6$. The complexity vanishes
at $\alpha_c$ which is expected to be the critical threshold for the
specific formula under study.}
 \label{cplx}
\end{figure}

In Fig. \ref{cplx} we report the data for the complexity of random
3-SAT formulae of size $N=10^6$ and $\alpha$ in the clustering
range. This has been obtained with the following procedure. One
generates first a 'random' 3-SAT formula with $N=10^6$ and $M=4.27\
10^6$, using a pseudo-random number generator.  The SP algorithm is
run on this formula, and the complexity is evaluated from
(\ref{sigma_def}). Then a new formula is generated from the previous
one by eliminating $10^4$ (pseudo-)randomly chosen clauses, and the
algorithm is run again, etc... It turns out that, for $N=10^6$, the
resulting curve is very `reproducible', in the sense that the
fluctuations of the curve from one random instance to the next are
small (typically below $1 \% $).

\subsection{Interpretation of the SID algorithm: categories of variables}

Once SP has reached convergence, we can compute the total biases $\{
W_i^{(\pm)}, W_i^{(0)} \}$. According to the previous discussion,
these numbers should be interpreted as giving the fraction of
constrained clusters where the variable $x_i$ is respectively (frozen
positive)/ (frozen negative)/ unconstrained.  Having computed these
weights, we may distinguish three reference types of variable nodes
(of course all the intermediate cases will also be present): the {\bf
under-constrained} ones with $W_i^{(0)} \sim 1$, the {\bf biased} ones
with either $W_i^{(+)} \sim 1$ or $W_i^{(-)} \sim 1$, and the {\bf
balanced} ones with $W_i^{(+)} \simeq W_i^{(-)}$ and $W_i^{(0)}$
small.

Fixing  a variable of each of these types produces different effects,
consistently with the interpretation of the surveys. The following
behaviours can be easily checked in numerical experiments: 

Fixing a biased variable does not alter the structure of the clusters and the
complexity changes smoothly (few constrained clusters are eliminated). 
This is the strategy used by the SID procedure. 

Fixing an under-constrained variable has no effect on the complexity.  

As expected, fixing a balanced variable  produces a decrease very
close to $\ln 2$ in the complexity.

\subsection{A summary of the main conjectures}

The whole interpretation relies on the existence, in certain `hard SAT'
regions of parameters (here in a window of $\alpha$ below $\alpha_c$),
of clusters of SAT assignments, which are very far apart (one cannot
reach a cluster from another one unless one flips a finite fraction of
the $N$ variables).  It would be very interesting to prove this
statement.

When there exist such clusters, some of them may be `constrained
clusters', in which some variables are constrained to take the same
value for all the assignments of the cluster. The introduction of the
don't-care state is an attempt at identifying all these clusters (in
which each cluster is characterized by a single point in the enlarged
assignment space including the don't-care state). SP, interpreted as
BP in this enlarged assignment space, is an attempt at obtaining a
statistical description of these constrained clusters.

\begin{figure}
\centering
\includegraphics[width=11.cm]{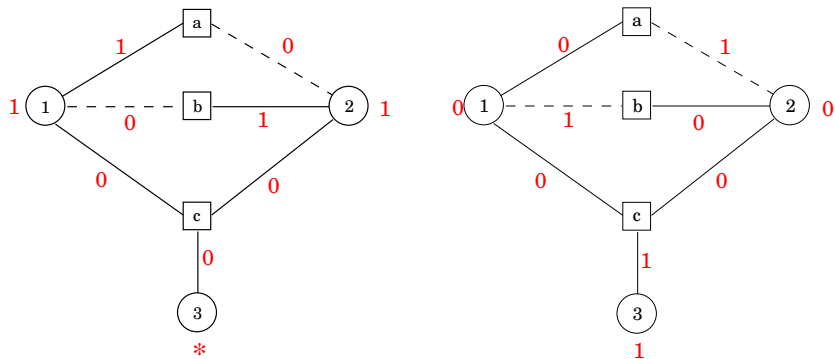}
\caption{ A simple SAT problem with three variables, two 2-clauses,
one 3-clause, and loops. Each figure gives one of the two possible
solutions of warning propagation. The number on each edge $(a,i)$ is
the warning $u_{a \to i}$, the number next to each variable node is
the corresponding generalized state of the variable in each of the two
'clusters'. The SP equations for this graph have an infinite number of
solutions, with $\eta_{a \to 1}= \eta_{b \to 2}=x$, $\eta_{a \to
2}=\eta_{b \to 1}=y$, $\eta_{c \to 1}= \eta_{c \to 2}=0$, and $\eta_{c
\to 3}=(1-x)^2 y^2/(1-x y )^2$ }
\label{fig_loopy}
\end{figure}

Two conjectures arise naturally from the heuristic statistical physics
approach and the numerical experiments on SP.  They should hold for
the random satisfiability problem, in the large $N$ limit, in some
hard SAT window of $\alpha$ just below $\alpha_c$ \footnote{After the
completion of this work, two groups have arrived to the conclusion
that the 'simple' clustering scenario described here should hold in
the region $\alpha \in [4.15,\alpha_c\simeq 4.2667]$ \cite{MMZ,MPR}}:
\begin{itemize}
\item 1) With probability one (on the set of initial messages),
SP converges to a unique set of fixed point messages.
\item 2) These fixed point messages
contain  the correct information about the constrained
clusters, and in particular the number of
constrained clusters can be computed as in (\ref{sigma_def}).

\end{itemize}
Note that for finite $N$, there are obvious counterexamples to these
conjectures, as shown in fig. \ref{fig_loopy}.

\section{Comments and perspectives}
\label{sect_concl}
When the solutions of a SAT (or more generally of a constrained
satisfaction problem) tend to cluster into well separated regions of
the assignment space, it often becomes difficult to find a global
solution because (at least in all local methods), different parts of
the problem tend to adopt locally optimal configurations corresponding
to different clusters, and these cannot be merged to find a global
solution. In SP we proceed in two steps.  First we define some
elementary messages which are warnings, characteristic of each
cluster, then we use as the main message the surveys of these
warnings. The warning that we used is a rather simplified object: it
states which variables are constrained and which are not constrained.
Because of this simplification the surveys can be handled easily,
which makes the algorithm rather fast.  As we saw in
sect.\ref{heuristic}, one might also aim at a finer description of
each cluster, where the elementary messages would give the probability
that a variable is in a given state ($\in \{0,1\}$), when considering
the set of all SAT configurations inside this given cluster. In this
case the survey will give the probability distribution of this
probability distribution, when one chooses a cluster randomly. This is
more cumbersome algorithmically, but there is no difficulty of
principle in developing this extension. One could also think of
generalizing further the messages (to probability distributions of
probability distributions of probability distributions), if some
problems have a structure of clusters within other clusters (this is
known to happen for instance in spin glasses \cite{MPV}), but the cost
in terms of computer resources necessary to implement it might become
prohibitive.

We have presented here a description of a new algorithmic strategy to
handle the SAT problem. This is a rather general strategy which can
also be applied in principle, to all Constraint Satisfaction Problems
\cite{coloring,MZ_pre}. At the moment the approach is very heuristic,
although it is based on a rather detailed conjecture concerning the
phase structure of random 3-SAT. The validity of a similar conjecture
has been checked exactly in the random XOR-SAT problem.  It would be
interesting to study this algorithmic strategy in its own,
independently from the random 3-SAT problem and the statistical
physics study. One can expect progress on SP to be made in the future
in various directions, among which: Rigorous results on convergence,
different ways of using the information contained in the surveys,
generalization of the algorithm to deal with complex graphs which are
not typical random graphs, use of the generalized SP with penalty to
provide UNSAT certificates. The generalization of SP to generic
Constraint Satisfaction Problems is discussed in a separate
publication \cite{BMWZ}.

\section*{Acknowledgments}
We are indebted to G. Parisi and M. Weigt for very fruitful
discussions. We also thank C. Borgs, J. Chayes, B. Hayes,
S. Kirkpatrick, S. Mertens, O. Martin, P. Purdom, M. Safra, B. Selman,
J. Yedidia for stimulating exchanges.  This work has been supported in
part by the European Community's Human Potential Programme under
contract HPRN-CT-2002-00319, 'STIPCO'.

\end{document}

%% file: table.tex
\begin{tabular}{|l|c|c|c|c|c|c|c|c|c|c|c|c|}
\hline 
$N=$&
\multicolumn{4}{c|}{$2.5\cdot10^{4}$}&
\multicolumn{4}{c|}{$5.0\cdot10^{4}$}&
\multicolumn{4}{c|}{$1.0\cdot10^{5}$}\tabularnewline
\hline
\hline 
$f$~\textbackslash{}~$\alpha$&
4.21&
4.22&
4.23&
4.24&
4.21&
4.22&
4.23&
4.24&
4.21&
4.22&
4.23&
4.24\tabularnewline
\hline
\hline 
4\%&
86\%&
66\%&
28\%&
8\%&
98\%&
84\%&
52\%&
22\%&
100\%&
100\%&
72\%&
22\%\tabularnewline
\hline 
2\%&
100\%&
86\%&
50\%&
22\%&
100\%&
98\%&
86\%&
48\%&
&
&
100\%&
68\%\tabularnewline
\hline 
1\%&
&
94\%&
78\%&
32\%&
&
100\%&
94\%&
64\%&
&
&
&
88\%\tabularnewline
\hline 
0.5\%&
&
98\%&
88\%&
50\%&
&
&
98\%&
66\%&
&
&
&
92\%\tabularnewline
\hline 
0.25\%&
&
100\%&
90\%&
60\%&
&
&
100\%&
78\%&
&
&
&
92\%\tabularnewline
\hline 
0.125\%&
&
&
94\%&
60\%&
&
&
&
84\%&
&
&
&
100\%\tabularnewline
\hline
\hline 
$<t>$&
1369&
2428&
4635&
7843&
1238&
1751&
3411&
8607&
1204&
1557&
2573&
7461\tabularnewline
\hline
\end{tabular}